\documentclass[%
preprint,
superscriptaddress,
showpacs,
aps,
prb,
]{revtex4-1}

\usepackage{graphicx}
\usepackage{dcolumn}
\usepackage{bm}
\usepackage{amsmath}
\usepackage{multirow}

\DeclareMathOperator{\erfc}{erfc}

\begin{document}


\title{Extended First-Principles Molecular Dynamics Method From Cold Materials to Hot Dense Plasmas}

\author{Shen Zhang}
\affiliation{HEDPS, Center for Applied Physics and Technology, Peking University, Beijing 100871, China}
\affiliation{College of Engineering,  Peking University, Beijing 100871, China}
\author{Hongwei Wang}
\affiliation{College of Engineering,  Peking University, Beijing 100871, China}
\author{Wei Kang}%
\email{weikang@pku.edu.cn}
\affiliation{HEDPS, Center for Applied Physics and Technology, Peking University, Beijing 100871, China}
\affiliation{College of Engineering,  Peking University, Beijing 100871, China}
\author{Ping Zhang}
\email{zhang\_ping@iapcm.ac.cn}
\affiliation{HEDPS, Center for Applied Physics and Technology, Peking University, Beijing 100871, China}
\affiliation{ LCP, Institute of Applied Physics and Computational Mathematics, Beijing 100088, P.R. China}
 \author{Xian-Tu He}
\email{xthe@iapcm.ac.cn}
\affiliation{HEDPS, Center for Applied Physics and Technology, Peking University, Beijing 100871, China}
\affiliation{ Institute of Applied Physics and Computational Mathematics, Beijing 100088, P.R. China}

\date{\today}

\begin{abstract}
An extended first-principles molecular dynamics (FPMD) method based on Kohn-Sham scheme is proposed to elevate the temperature limit of the FPMD method in the calculation of dense plasmas. The extended method treats the wave functions of high energy electrons as plane waves analytically, and thus expands the application of the FPMD method to the region of hot dense plasmas without suffering from the formidable computational costs. 
In addition, the extended method inherits the high accuracy of the Kohn-Sham scheme and keeps the information of electronic structures.  This gives an edge to the extended method in the calculation of the lowering of ionization potential, X-ray absorption/emission spectra, opacity, and high-Z dense plasmas, which are of particular interest to astrophysics, inertial confinement fusion engineering, and laboratory astrophysics.
\end{abstract}

\maketitle
\thispagestyle{headings}


\section{Introduction}
Properties of dense plasmas, including their equation of state (EOS), opacity, and X-ray absorption spectra, etc., are substantial subjects of the emerging field of high energy density physics\cite{drake2006} and laboratory astrophysics.\cite{Remington1999,remington2006experimental} They are not only key parameters in the design of inertial confinement fusion (ICF)  targets,\cite{atzeni2004} but also of fundamental interest to the understanding of the evolution and structure of celestial bodies such as the sun, giant planets, and brown dwarfs.\cite{Guillot1999}    

After decades-long combined efforts from both theoretical and experimental parts,\cite{wdmbook2014,remington2006experimental} it has been recognized that an adequate quantum-mechanical description of electrons is essential for theoretical calculations to afford a satisfactory prediction to the property of dense plasmas. There are a variety of first-principles approaches\cite{wdmbook2014, dai2010QLMD, sjostrom2014} on different approximation levels to address this issue. Among them, the first-principles molecular dynamics (FPMD),\cite{surh2001, Huser2005, galli2000, lenosky2000} orbital-free molecular dynamics (OFMD),\cite{lambert2006, wang2011, dt} and path-integral Monte Carlo (PIMC)\cite{CeperleyPIMC, PIMCelectrongas,hu2011PIMC,militzer2009PIMC} methods are most popularly employed.

The FPMD method combines a quantum-mechanical treatment to the fast moving electrons and a classical treatment to the slowly varying ionic part.\cite{Barnett1991, kresse1993BOMD} Electrons are described in the Kohn-Sham scheme\cite{hohenberg1964,kohn1965} with  many-body effect of  electrons accounted for by the finite-temperature density functional theory(FT-DFT).\cite{mermin1965FTDFT} The ions are treated classically according to Newton's equation of motion with the force acted upon them determined by the Hellamnn-Feynman theorem or its finite-temperature generalization\cite{kittel_tp} with the  Born-Oppenheimer approximation assumed.\cite{Barnett1991}
The FPMD method is very successful when the temperature of plasmas is relatively low, usually not above the warm dense region.\cite{wdmbook2014} When temperature further increases, the electronic states of non-negligible contribution arrive at a high energy on the order of tens of $T$, which, combined with the 3/2 scaling of the number of electronic states with respect to the energy, makes the FPMD method practically inapplicable at high temperature.

A remedy to the prohibitive computational costs at high temperature is provided by the OFMD method.\cite{lambert2006} It gives up the explicit description of electronic orbitals, i.e., without information on wave functions and eigen-energies of electronic states, which are fundamental concepts of the Kohn-Sham scheme and the base for the calculation of X-ray absorption and opacity.\cite{recoules2009, zhang2015arxiv} Instead, the method adopts the original idea of density functional theory, formulating the free energy as a functional of electron charge density.\cite{karasiev2014,lambert2006} Usually, the Thomas-Fermi approximation is applied to the kinetic part. It has a much improved computational efficiency but at the expense of being less accurate at low temperature.\cite{wdmbook2014}

The PIMC method\cite{CeperleyPIMC} deals directly with the density matrices in quantum statistical formulas, treating both ions and electrons as paths equally.\cite{wdmbook2014, hu2011PIMC} By expressing the density matrices as path integrals, physical properties are calculated with the help of the Monte Carlo sampling. Limited by the ``Fermion sign problem'' of electrons, the method is trusted as an accurate quantum mechanical solution to plasmas at temperature much higher than the Fermi temperature, where the effect of degeneracy is not important.\cite{wdmbook2014, militzer2009PIMC}

It is desirable to have a systematic method practically covering the entire temperature range, i.e., from cold materials to hot dense plasmas, at the least expense of accuracy and computational costs. In addition, it will be helpful if the method also provides the information of electronic structures, which are important physical quantities diagnosed by the optical and X-ray approaches in laboratory astrophysics\cite{Remington1999,remington2006experimental} and ICF.\cite{atzeni2004}  

We show that this can be done in the framework of Kohn-Sham scheme by approximating the electronic orbitals at high energy as plane waves analytically. With minor modifications, the FPMD method is readily extended to the high-temperature region, i.e., the region of hot dense plasmas, without suffering from the explosively growing computational costs, and still retains the accuracy and advantage of the FPMD method. The extended method, hereinafter referred as ext-FPMD,  not only expands the temperature range that the FPMD method can deal with, 
but can also be used to illustrate subtle features of hot dense plasmas closely related to electronic structures, e.g., the effect of electronic shells upon compression, the X-ray absorption spectra, and the lowering of ionization potential.

The rest of the article is organized as follows. The theory of the ext-FPMD method is developed in Sec.~II, and implementation details in plane-wave bases are discussed in Sec.~III. Comparisons of the ext-FPMD method with other first-principles methods are presented in Sec.~IV,  using deuterium (D), $^6$LiD, helium (He), and aluminum (Al) as illustrating examples. The article is concluded in Sec.~V with a short summary. All formulas are presented in the atomic units with the Boltzmann constant $k_B$ set as 1.

\section{Development of the theory}
\subsection{Plane-wave approximation at high energy}

\begin{figure}
\centering
\includegraphics[scale=0.3]{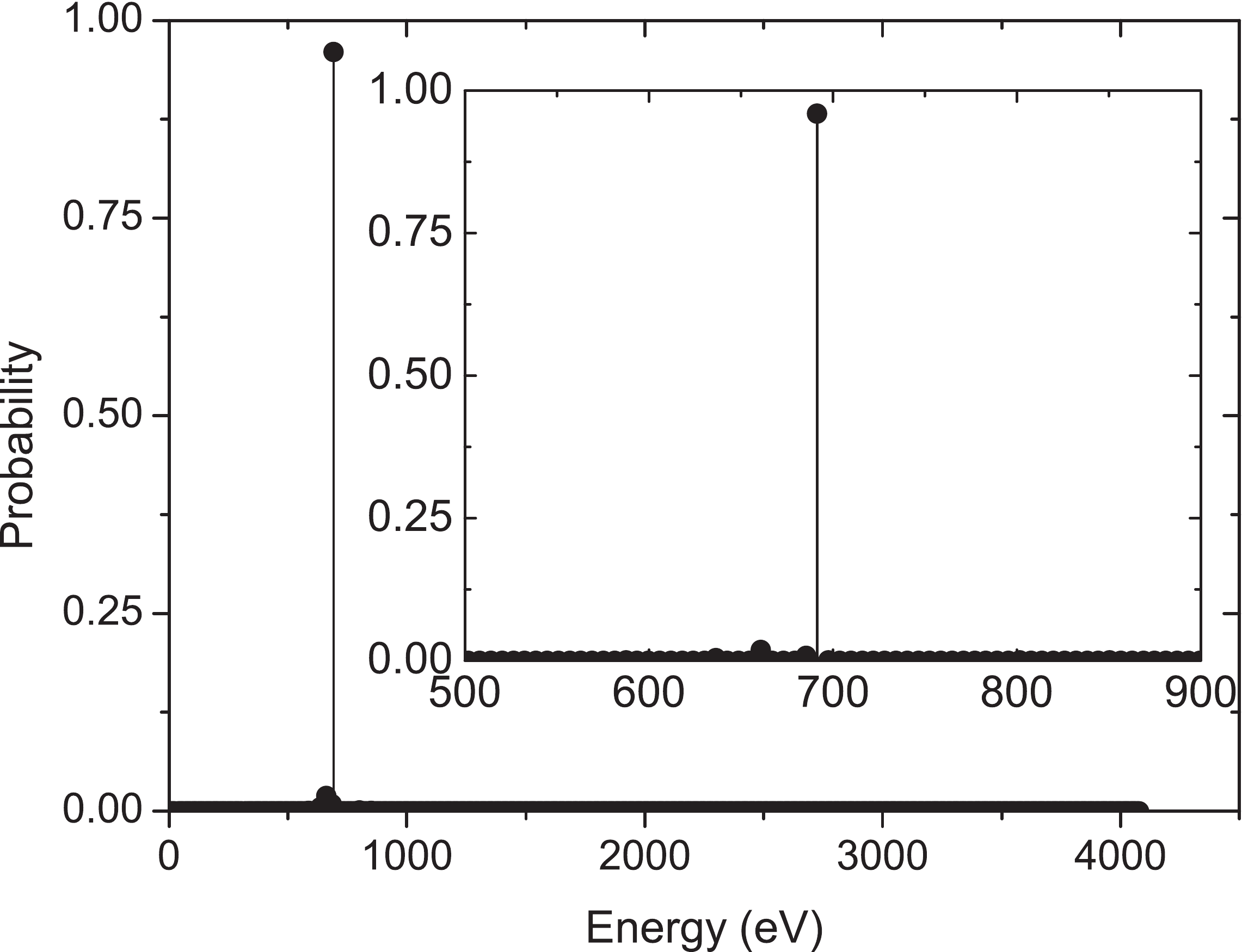} 
\caption{\label{fig_fourier}(Color online) Plane-wave decomposition of a  Kohn-Sham wave function associated with ${\bf k}$ = (7/6 1/6 1/6) and $|{\bf k}|$ = 7.1295 in warm dense Al generated by high power laser,\cite{vinko2014} displayed as a function of plane-wave energy. The corresponding plasma state is $\rho$ = 2.7 g/cm$^3$, $T_e$ = 13.6 eV, and  Al ions are frozen in the FCC lattice. Inset: the energy scale is zoomed in  around 700 eV to highlight the single peak structure of the orbital. }
\end{figure}
 
From the point of view of  independent electron approximation,\cite{martin2004Book} the motion of high energy electrons in plasmas is well illustrated by the scattering of electrons from the screened Coulomb potential $V(r) \propto  \alpha/r \exp(-\kappa r)$, where $r$ is the distance between electrons and ions, $\alpha$ is the effective charge, and $\kappa$ is the screening parameter. The $V(r)$ here represents the most typical electron-ion interaction in plasmas.\cite{drake2006} Following the standard scattering theory, the wave function $\psi$ at energy $\epsilon$ can be approximately written as $\psi \sim \exp(ikz)+f(\theta)\exp(ikz)/r$, where the $z$ axis is taken as the propagating direction of the plane wave, $\theta$ is the azimuthal angle with respect to the $z$ axis, and $k=\sqrt{2\epsilon}$ is the wave vector of the plane wave. The deviation of $\psi$ from a plane wave is accounted for by the total scattering cross-section $\sigma_t =2\pi \int d\theta \sin\theta |f(\theta)|^2$, which is $\frac{\alpha^2}{ \kappa^2}\frac{16\pi}{8\epsilon+\kappa^2}$ as calculated from the Born approximation.\cite{landau_qm} It shows that the total deviation of $\psi$ from a plane wave decreases as $\epsilon^{-1}$ when the energy of an electron is high enough. In another word,  high energy electrons in plasmas behave like free electrons.

This resemblance of free electrons can also be displayed numerically in real materials. For example, Fig.~\ref{fig_fourier} shows the plane-wave decomposition of the Kohn-Sham electronic orbital of Al corresponding to ${\bf k}$ = (7/6 1/6 1/6) and $|{\bf k}|$ = 7.1295 in a warm dense state generated by high power laser,\cite{vinko2014} at which the electrons is heated to a temperature of 13.6 eV, while the Al ions retain unmoved at the lattice position. The uncompressed Al has a face-centered cubic (FCC) lattice with  lattice parameter $a=$ 4.05 {\AA} 
at room temperature. The result is calculated using the FT-DFT method with the local density approximation (LDA) to the exchange-correlation interaction. An plane-wave cutoff of 300 Ry and an unshifted 6$\times$6$\times$6 Monkhorst-Pack (M-P) k-point mesh\cite{Monkhorst1976} is adopted in the calculation. The pseudopotential of Al takes the projector augmented wave (PAW) format\cite{blochl1994} with $2s$ $2p$ $3s$ $3p$ electrons explicitly included as valence electrons. The decomposition in Fig.~\ref{fig_fourier} is presented as a function with respect to the energy of the plane waves. The single peak at $E$= 691.575 eV shows that the orbital is mainly composed of  plane waves (actually two plane waves because of the time-reversal symmetry) at the given energy. 

There are two immediate consequences following the plane wave approximation at high energy, which form the foundation of the ext-FPMD method. One is that the charge density of high energy electrons is uniformly distributed in the space. This is evident from the approximation, and we shall further show in Sec.~IV that the deviation is small compared with a direct FT-DFT calculation.
Another consequence is that the spatial undulation of the potential field experienced by the high energy electrons can be neglected. This implies that the potential field can be treated as a  constant background effectively. Accordingly, the density of state (DOS) $D(\epsilon)$ of high energy electrons is expressed as 
\begin{equation}
D(\epsilon) = \frac{\sqrt{2}\Omega}{\pi^2}\sqrt{\epsilon-U_0} \label{eq_dos}
\end{equation}
where $\epsilon$ is the energy of an electronic state, $\Omega$ is the volume of the calculation cell, and $U_0$ is the constant effective potential energy.  Since $\epsilon-U_0$ is the kinetic energy $E_K$ of an electron, Eq.(\ref{eq_dos}) is exactly the DOS formula of a free electron in a constant potential field.

\begin{figure}
\centering
\includegraphics[scale=0.3]{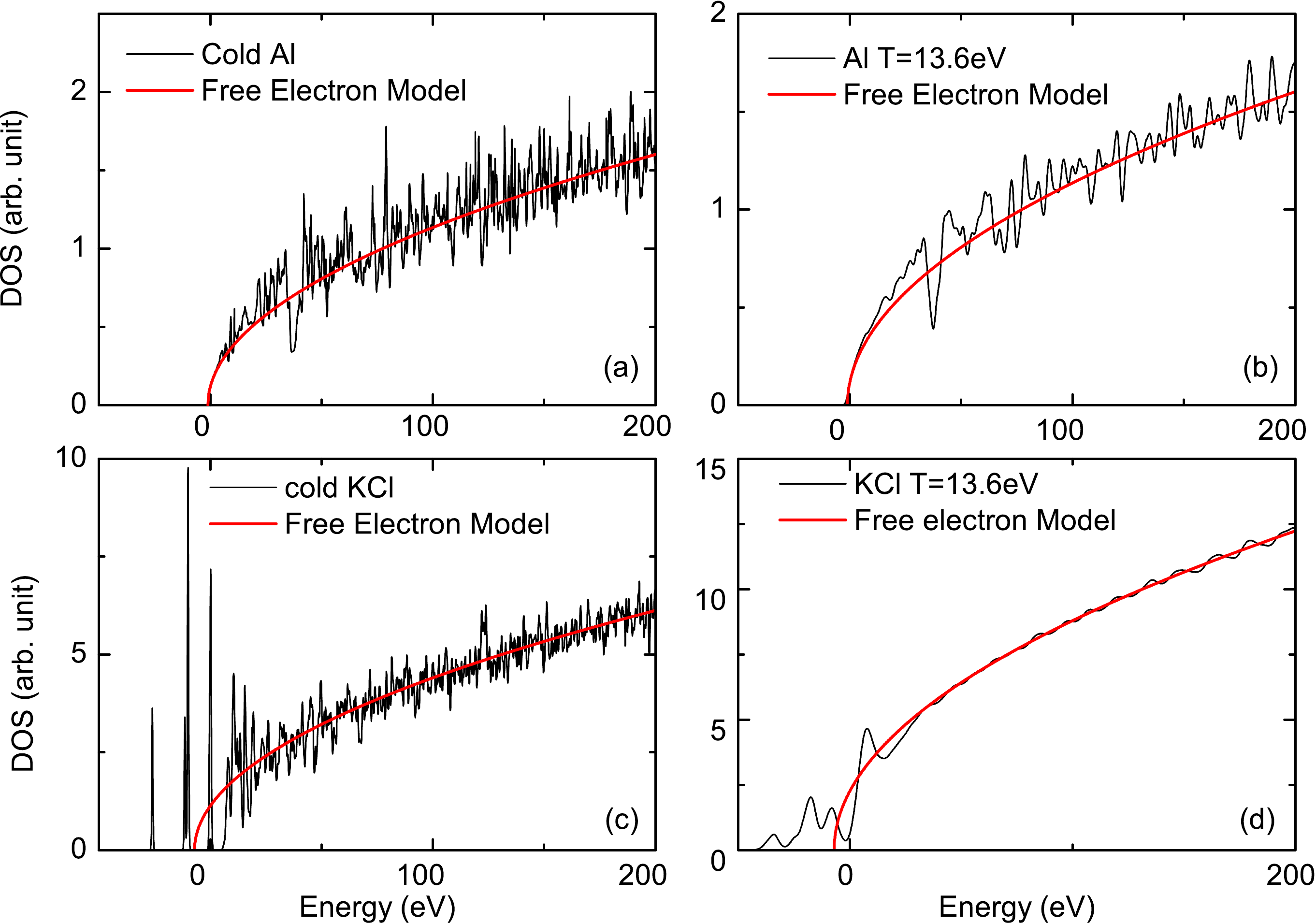} 
\caption{\label{fig_dos}(Color online) DOS predicted by the free electron model in Eq.(\ref{eq_dos}) compared with DOS calculated by the FPMD method under typical plasma conditions. (a) Cold electrons in a FCC Al lattice at $\rho$ = 2.7 g/cm$^3$; (b) hot electrons at $T_e$ = 13.6 eV in the same Al lattice as in (a); (c) cold electrons in a KCl FCC crystal at $\rho$ = 1.984 g/cm$^3$; and (d) fully developed plasmas of KCl at $T$ = 13.6 eV and $\rho$ = 1.984 g/cm$^3$, calculated with 4 atoms. The red solid lines represent the prediction of free electron model, and the black solid curves are DOS of FPMD calculation. Except for the low energy part, all the DOS are perfectly described by the free electron model.}
\end{figure}

The validity of Eq.(\ref{eq_dos}) is extensively examined under a variety of conditions. In Fig.~\ref{fig_dos}, we display the comparison of Eq.(\ref{eq_dos}) with the DOS  calculated with the FPMD method under four typical conditions. 
Fig.~\ref{fig_dos}(a) is the DOS of cold electrons , i.e., $T_e$ = 0 K, in a periodic FCC lattice of Al at $\rho$ = 2.7 g/cm$^3$, representing the condition of cold metallic crystals. 
Fig.~\ref{fig_dos}(b) is the DOS of hot electrons at $T_e$ = 13.6 eV in a periodic FCC lattice of Al the same as that in Fig.~\ref{fig_dos}(a). It represents the dense plasmas generated by a high power fast laser through isochoric heating processes.\cite{vinko2014}  The condition of cold electrons in an insulating crystal is illustrated by potassium chloride (KCl) in Fig.~\ref{fig_dos}(c), calculated in a FCC lattice at $\rho$ = 1.984 g/cm$^3$. The condition of full-developed dense plasmas is displayed in Fig.~\ref{fig_dos}(d), calculated with KCl at $T_e$ = $T_i$ = 13.6 eV and $\rho$ = 1.984 g/cm$^3$. All the DOS calculated with the FPMD methods are displayed as black curves, while the numerical results of Eq.(\ref{eq_dos}) are drawn as red lines. Fig.~\ref{fig_dos} shows that  Eq.(\ref{eq_dos}) provides a fairly good description to the DOS at high energy for all the conditions. The fluctuation of the DOS in Fig.~\ref{fig_dos}(a)-(c) is the result of resonant scatterings from the crystalline lattice,\cite{landau_qm} which almost disappears in Fig.~\ref{fig_dos}(d) when the ion lattice is destroyed in fully developed plasmas.  

It is also noticed from Fig.~\ref{fig_dos} that the $U_0$ actually represents a constant shift of free electron DOS on the energy scale, typically on the order of 10 eV. Once the $U_0$ is determined, one has a good approximation to the DOS at high energy. This can be done by two different approaches. First, since $U_0$ is the only unknown parameter in Eq.(\ref{eq_dos}), it can be determined by fitting Eq.(\ref{eq_dos}) to the DOS obtained from the FPMD calculation. Second, it can be calculated as the average of $\epsilon-E_K$ at high energy. 

\subsection{FT-DFT With Plane Wave Approximation at High Energy}

One can now incorporate the plane wave approximation at high energy into the FT-DFT. Formulas are presented in a spin-averaged form assuming the plasma is spin-unpolarized.  Different parts from the original FT-DFT formulas are marked with a superscript $\mathcal R$.

Within the Kohn-Sham scheme, the Mermin's grand potential functional\cite{mermin1965FTDFT} for electrons is formally written as
$
\Xi=E-TS-\mu N,
$
where $E$ is the total energy, $T$ is the temperature of electrons,  $S$  is the entropy, $\mu$ is the chemical potential, and $N$ is the number of electrons in the system. $E$ and $S$ are explicitly expressed as
\begin{equation*}
\begin{aligned}
E=&-\sum_{i=1}^{\infty}f(\epsilon_i)\left<\psi_i\right|\nabla^2\left|\psi_i\right> \\
             &  + E^{h}[n]+E^{xc}[n]+\int d{\bf r} V^{ei}({\bf r}) n({\bf r}),
\end{aligned}
\end{equation*}
and
\begin{equation*}
S = -2\sum_{i=1}^{\infty}\left\{f(\epsilon_i) \ln f(\epsilon_i) -\left[1-f(\epsilon_i)\right] \ln \left[1-f(\epsilon_i)\right]\right\},
\end{equation*} 
respectively, where $i$ is the index of energy eigenstates, $E^{h}$ is the Hartree energy, $E^{ex}$ is the exchange-correlation energy, $V^{ei}$ is the ionic potential experienced by the electrons,  
and $n({\bf r})$ is the charge density of electrons. $f(\epsilon_i)=1/[\exp(\epsilon_i-\mu)/T+1]$ represents the Fermi-Dirac distribution, which makes the grand potential $\Xi$ reaches its minimum at the constant $N$ constraint. 
The energies $\epsilon_i$, the wave functions $\psi_i({\bf r})$, the chemical potential $\mu$, and the charge density $n({\bf r})$ 
are self-consistently determined from the variational Kohn-Sham equation 
\begin{equation}\label{eq_ks}
\left[-\frac{1}{2}\nabla^2 +V^{h}[n]+V^{ex}[n]+V^{ei}({\bf r})\right]\psi_i({\bf r})=\epsilon_i\psi_i({\bf r}), 
\end{equation}
with   
\begin{equation}\label{eq_n}
n({\bf r})=2\sum_{i=1}^{\infty}f(\epsilon_i)|\psi_i({\bf r})|^2, 
\end{equation}
and $\mu$ determined by the charge conservation equation 
\begin{equation}
N=2\sum_{i=1}^{\infty}f(\epsilon_i)=2\sum_{i=1}^{\infty}\frac{1}{e^{(\epsilon_i-\mu)/T}+1}. \label{eq_mu}
\end{equation}

The plane wave approximation amounts to treating $V^{h}[n]+V^{ex}[n]+V^{ei}({\bf r})$ as a constant potential $V_0$ for $i \gg1$, which reduces Eq.(\ref{eq_ks}) to a free-electron Kohn-Sham equation that has a simple analytical solution. Here $V_0$ is just $-U_0$ in Eq.(\ref{eq_dos}), because the charge of an electron is $-1$ in the atomic units. The $\epsilon_i$ and $\psi_i({\bf r})$ at high energy can now be substituted by the analytical solutions.

It turns out that only minor revisions in Eq.(\ref{eq_n}) and Eq.(\ref{eq_mu}) are needed to incorporate the approximation into the self-consistent Kohn-Sham equation. With the approximation, Eq.(\ref{eq_n}) is revised as 
\begin{equation}\label{eq_n_rev}
n^{^\mathcal R}({\bf r})=2\sum_{i=1}^{N_c}f(\epsilon_i)|\psi_i({\bf r})|^2+\frac{1}{\Omega}\int_{E_c}^{\infty}d\epsilon f(\epsilon)D(\epsilon), 
\end{equation}
and Eq.(\ref{eq_mu}) turns to be 
\begin{equation}\label{eq_mu_rev}
N^{^\mathcal R}=2\sum_{i=1}^{N_c}\frac{1}{e^{(\epsilon_i-\mu)/T}+1}
    + \int_{E_c}^{\infty}d\epsilon f(\epsilon)D(\epsilon), 
\end{equation}
given $E_c$ the lowest energy bound of the plane wave approximation, which corresponds to the energy of the $N_c$-th electronic state. Here $U_0$ in the expression of  $D(\epsilon)$ is determined as the average of the potential energy $\epsilon_i+\frac{1}{2}\left<\psi_i\right|\nabla^2\left|\psi_i\right>$ near $E_c$, which is equivalent to fitting Eq.(\ref{eq_dos}) with the FT-DFT DOS.

Accordingly, the total energy $E$ has extra corrections from $n^{^\mathcal R}$ and kinetic energy of  high-energy electrons. With these corrections, $E$ is revised as 
\begin{equation}\label{eq_e_rev}
\begin{aligned}
E^{^\mathcal R}&=-\sum_{i=1}^{N_c}f(\epsilon_i)\left<\psi_i|\nabla^2 |\psi_i\right> \\
            +& \int_{E_c}^{\infty}d\epsilon f(\epsilon)D(\epsilon)(\epsilon-U_0)  \\
             +&E^{h}[n^{^\mathcal R}]+E^{xc}[n^{^\mathcal R}]+\int d{\bf r} V^{ei}({\bf r}) n^{^\mathcal R}({\bf r}).
\end{aligned}
\end{equation}
Similarly, the entropy is revised as 
\begin{equation}\label{eq_s_rev}
\begin{aligned}
S&^{^\mathcal R} = -2\sum_{i=1}^{N_c}\left\{ f(\epsilon_i) \ln f(\epsilon_i) 
    -\left[1-f(\epsilon_i)\right] \ln \left[1-f(\epsilon_i)\right] \right\} \\
    & -\int_{E_c}^{\infty}d\epsilon D(\epsilon)\left\{ f(\epsilon)\ln f(\epsilon)+\left[1-f(\epsilon)\right]\ln \left[1-f(\epsilon)\right] \right\},
\end{aligned}
\end{equation} 
by taking the contribution of high energy electrons into account.

\begin{figure}
\centering
\includegraphics[scale=0.35]{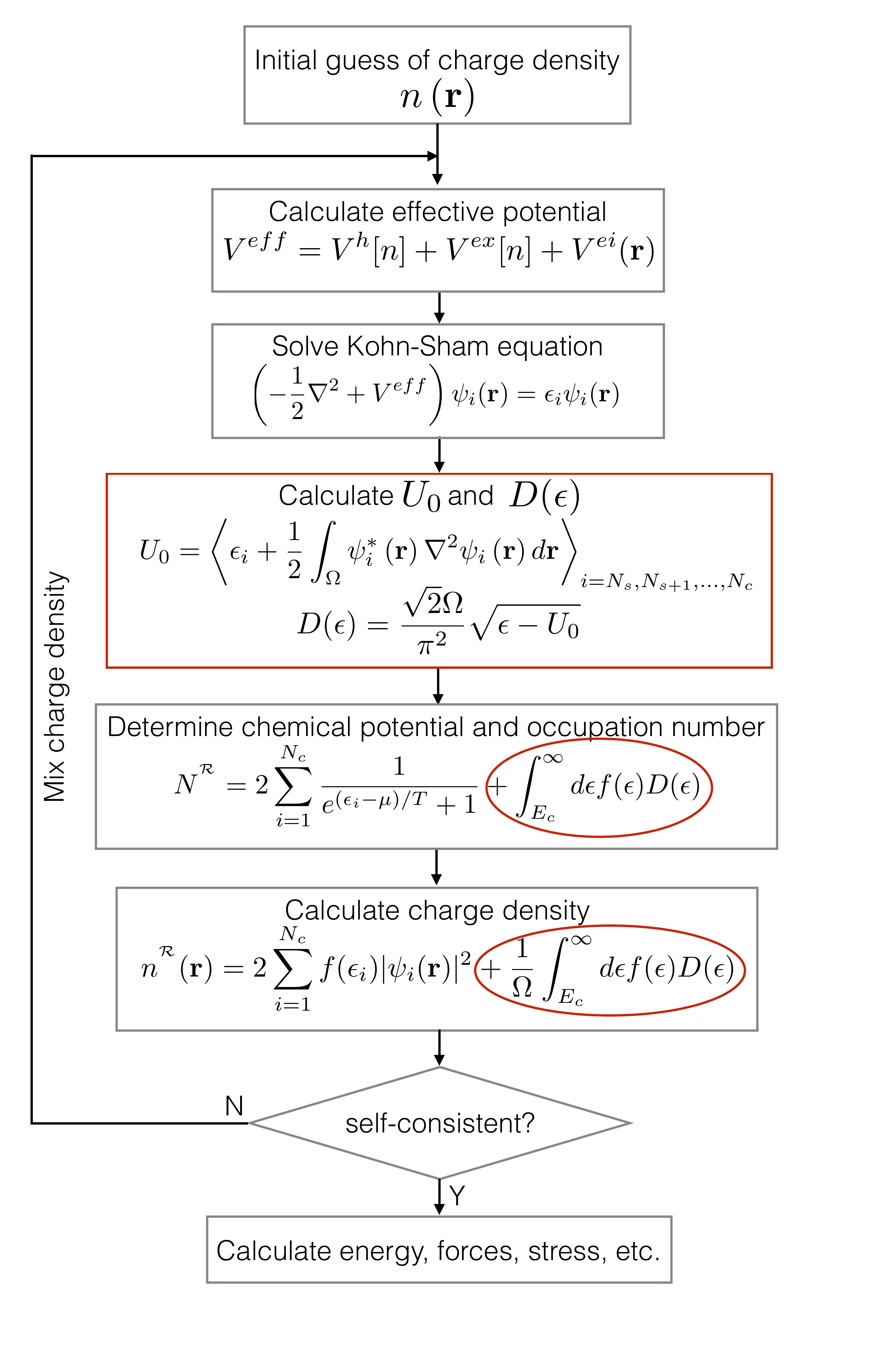} 
\caption{\label{fig_iter}(Color online) Flow chart of the self-consistent iterating process in the extended FT-DFT method with plane-wave approximation at high energy, where $N_s$ denotes the starting index of the electronic orbital used to calculate $U_0$.}
\end{figure}
With all revisions wrapped up together, the FT-DFT with plane-wave approximation at high energy is then composed of five equations, which are the  Kohn-Sham equation in Eq.(\ref{eq_ks}),  the revised charge density expression in Eq.(\ref{eq_n_rev}), the revised charge conservation of Eq.(\ref{eq_mu_rev}) to determine $\mu$, together with total energy and entropy calculated by Eq.(\ref{eq_e_rev}) and Eq.(\ref{eq_s_rev}) respectively. The flow chart of the iterating process is summarized in Fig.~\ref{fig_iter}, where $N_S$ denotes the starting index of the electronic orbital used to determine $U_0$, which is smaller than the onset of plane wave approximation $N_c$. Revised parts are marked with red frames.

\subsection{Forces and Stresses -- Coupling with Born-Oppenheimer Molecular Dynamics}

In the calculation of plasmas, the FT-DFT has to couple with the dynamics of ions, usually through the Born-Oppenheimer (BO) molecular dynamics (MD) method in recent works.\cite{surh2001, Huser2005, galli2000, lenosky2000} Since the motion of ions are treated classically in the BOMD method, it is necessary to extract the force on ions, which is calculated by the Hellman-Feynman force theorem or its finite-temperature generalization\cite{kittel_tp} as ${\bf F}_I=-(\partial E/\partial {\bf R}_I)_S$ after the Kohn-Sham equation is solved self-consistently. Here, the subscript $S$ indicates the partial derivative is carried out  isentropically.  Its FT-DFT version 
\begin{equation}\label{eq_f}
{\bf F}_I= - \int d{\bf r} n({\bf r})\frac{\partial V^{ei}({\bf r}; \{{\bf R}_I\})}{\partial {\bf R}_I}-{\bf F}_{I}^{ii},
\end{equation}
only depends on the ion-ion interaction as well as the electron-ion interaction.
Here $\{{\bf R}_I\}$ represents the the positions of all ions. ${\bf F}_I$ is the force on the I-th ion, and ${\bf F}_{I}^{ii}$ is the force imposed by other ions. 

In principle, one just need to use the revised $n^{^\mathcal R}({\bf r})$ in Eq.(\ref{eq_n_rev}) to calculate the force in Eq.(\ref{eq_f}) when the plane wave approximation at high energy is applied. It assumes $\hat{V}^{ei}$ is local, i.e., the operator $\hat{V}^{ei}({\bf r}, {\bf r}')$ has a diagonal form of $\hat{V}^{ei}({\bf r}, {\bf r}')=\delta({\bf r}-{\bf r}')V^{ei}({\bf r})$. 
This is the case when all electrons are explicitly considered. Then, $V^{ei}({\bf r})$ is just the superposition of Coulomb potentials of ions. However, when the pseudopotential technique\cite{martin2004Book} is used to describe the effective interaction between ions and electrons, the non-locality of the $\hat{V}^{ei}$ may bring about some subtleties. It is not a concern for the calculation of forces, because high energy electrons, now approximated as plane waves, are almost not scattered by ions.

The pressure $P$, which is a principal parameter in the EOS of plasmas, is derived from the stress tensor $\sigma_{\alpha \beta}=\Omega^{-1}(\partial E/\partial \varepsilon_{\alpha\beta})_{S}$ as $P=-(1/3)\sum_{\alpha}\sigma_{\alpha \alpha}$, where $\varepsilon_{\alpha\beta}$ is the symmetric strain tensor, and the subscript $S$ indicates that the $\sigma_{\alpha \beta}$ is calculated under isentropic conditions.  The stress tensor $\sigma_{\alpha\beta}$ can formally break into several terms as
\begin{equation}\label{eq_stress}
\sigma_{\alpha\beta}=\sigma_{\alpha\beta}^{ii}+\sigma_{\alpha\beta}^k+\sigma_{\alpha\beta}^{ei}+\sigma_{\alpha\beta}^h+\sigma_{\alpha\beta}^{xc},
\end{equation}
where the subscripts $\alpha$ and $\beta$ denote the components of the stress tensor. $\sigma_{\alpha\beta}^{ii}$ is the contribution of ions, $\sigma_{\alpha\beta}^k$ is the contribution of kinetic energy of electrons, $\sigma_{\alpha\beta}^{ei}$ comes from ion-electron interactions, $\sigma_{\alpha\beta}^h$ comes from the Hartree energy of electrons,  and $\sigma_{\alpha\beta}^{xc}$ is from the electronic exchange-correlation interaction. 

In the framework of BOMD, $\sigma_{\alpha\beta}^{ii}$ is calculated classically, and one has
$$
\sigma_{\alpha\beta}^{ii} =-\sum_I\frac{P_{I,\alpha}P_{I,\beta}}{m_i}-\sum_I R_{I,\alpha}F_{I,\beta},
$$
according to Ref.~\onlinecite{allen1987}, where ${\bf P}_I$ is the momentum of the I-th ion. The formula does not change when the plane wave approximation at high energy is applied, because no electronic properties are explicitly presented. 

The original form of $\sigma_{\alpha\beta}^k$ in FT-DFT is\cite{nielsen1985stress} 
$$
\sigma_{\alpha\beta}^k = 2\sum_{i=1}^{\infty}f(\epsilon_i)\left<\psi_i|{\nabla}_{i,\alpha}{\nabla}_{i,\beta}|\psi_i\right>,
$$
where $\nabla_{i,\alpha}$ is the abbreviation of ${\partial}/{\partial r_{i,\alpha}}$. With the plane-wave approximation at high energy, it is revised as
$$
\begin{aligned}
(\sigma_{\alpha\beta}^{k})^{^\mathcal R} &= 2\sum_{i=1}^{N_c}f(\epsilon_i)\left<\psi_i|{\nabla}_{i,\alpha}{\nabla}_{i,\beta}|\psi_i\right>\\
                                   &-\frac{2\delta_{\alpha\beta}}{3}\int_{E_c}^{\infty}d\epsilon D(\epsilon)f(\epsilon)(\epsilon-U_0),
\end{aligned}
$$
where $\delta_{\alpha\beta}$ is the Kronecker $\delta$ function, and $\epsilon-U_0$ is the kinetic energy of  high energy electrons.  
The non-diagonal contribution of high energy electrons to $\sigma_{\alpha\beta}^{^{\mathcal R}}$ vanishes because electrons move in all directions with equal probability. This is evident when the integral $\int_{\epsilon>E_c}^{\infty}d{\bf p}(\epsilon) D(\epsilon)f(\epsilon)p_{\alpha}p_{\beta}$  is performed in the spherical coordinates $(p,\theta,\phi)$ of momentum $\bf p$ space, where $p_x=(\epsilon-U_0)\sin\theta \cos\phi$, $p_y=(\epsilon-U_0)\sin\theta\sin\phi$, and $p_z=(\epsilon-U_0)\cos\theta$. 

With the plane wave approximation at high energy, $\sigma^{h}_{\alpha\beta}$ is calculated as 
$$
\sigma^h_{\alpha\beta}=-\frac{1}{2}\iint d{\bf r} d{\bf r}' n^{^{\mathcal R}}({\bf r}) n^{^{\mathcal R}}({\bf r}') \frac{({\bf r}-{\bf r}')_{\alpha}({\bf r}-{\bf r}')_{\beta}}{|{\bf r}-{\bf r}'|^3},
$$
 which has the same form as that in the original FT-DFT formula\cite{nielsen1985stress} but with $n({\bf r})$ and $n({\bf r}')$ substituted with those calculated through Eq.(\ref{eq_n_rev}). 

The revised $\sigma_{\alpha\beta}^{ei}$ can be expressed formally as 
$$
\begin{aligned}
\sigma_{\alpha\beta}^{ei}=&-\frac{1}{2}\sum_I \int d{\bf r} n^{^{\mathcal R}}({\bf r}) \left. \frac{dv^C(\xi)}{d\xi}\right|_{\xi=|{\bf r }-{\bf R}_I|} \\
                                          & \times \frac{({\bf r}-{\bf R}_I)_{\alpha}({\bf r}-{\bf R}_I)_{\beta}}{|{\bf r}-{\bf R}_I|}
\end{aligned}
$$
with $v^C$ the Coulomb potential of ions, and $n^{^{\mathcal R}}({\bf r})$ calculated via Eq.(\ref{eq_n_rev}). The employment of pseudopotential may bring about extra corrections to $\sigma_{\alpha\beta}^{ei}$. However, as we shall discuss in the next section, these corrections are generally small for high energy electrons and will not explicitly considered in the extended FT-DFT method. 

The expression of revised $\sigma^{xc}_{\alpha\beta}$ depends on the choice of exchange-correlation functionals and can be complex when orbital-dependent functionals are employed. For the LDA and generalized gradient approximation (GGA) 
exchange-correlation functionals commonly used in plasma calculations, the expression of $\sigma_{\alpha\beta}^{xc}$  has the same form as that in the FT-DFT.\cite{martin2004Book} For LDA, 
\begin{equation}\label{eq_sigma_xc_lda}
\sigma_{\alpha\beta}^{xc, LDA}=\delta_{\alpha\beta}\int d{\bf r} n^{^{\mathcal R}}({\bf r}) \left\{ \epsilon^{xc}[n^{^{\mathcal R}}({\bf r})]-v^{xc}[n^{^{\mathcal R}}({\bf r})]\right\},
\end{equation}
where $\epsilon^{xc}$ is the exchange-correlation energy density defined by $E^{xc}[n({\bf r})]=\int d{\bf r}n({\bf r}) \epsilon^{xc}[n({\bf r})]$, and $v^{xc}=d(n \epsilon^{xc})/dn$. For GGA functionals in the form of $E^{xc}[n({\bf r})]=\int d{\bf r}n({\bf r}) \epsilon^{xc}[n({\bf r}), \nabla n({\bf r})]$,  an extra correction term 
$$
\sigma_{\alpha\beta}^{xc, GGA}= \int d{\bf r} \nabla_\alpha n^{^{\mathcal R}}({\bf r}) \nabla_\beta \left\{n^{^{\mathcal R}}({\bf r}) \epsilon^{xc}[n^{^{\mathcal R}}({\bf r}), \nabla n^{^{\mathcal R}}({\bf r})]\right\}
$$
is presented for the density gradient.

\section{Implementation in Plane-Wave Bases with Pseudopotentials}
It is straightforward to implement the approximation in plane-wave bases. Our implementation is based on the \texttt{Quantum-Espresso} package,\cite{pwscf} and all calculations in this work are carried out  using the code. 

\subsection{Technical details on the extended FT-DFT}
There are several technical details needed to be noted for the usage of pseudoptentials, where an angular-momentum dependent nonlocal potential operator ${\hat V}^{NL}_l$ may be involved to represent the effective ion-electron interaction. Here, $l$ stands for the angular quantum number. 

The residue error caused by the nonlocal operator ${\hat V}^{NL}_l$ varies with the type of pseudopotentials. The calculations using PAW pseudopotentials are essentially all-electron calculations.\cite{blochl1994} The non-locality of ${\hat V}^{NL}_l$ (operated on the smooth pseudo-wave functions) is canceled  by the non-locality of the on-site correction in the augmentation region, and eventually the locality of the Coulomb potential is recovered approximately. The accuracy of the pseudopontential relies on the details of the implementation and the number of scattering channels.  So, in principle, there is no error brought about by the non-local ${\hat V}^{NL}_l$ in the PAW pseudopoentials when they are properly constructed. 

But the situation is different when norm-conserving pseudopotentials (NCPP) are used, where the non-local ${\hat V}^{NL}_l$ is the intrinsic feature of the method. There are additional residue errors caused by the ${\hat V}^{NL}_l$, which makes the calculation less accurate than those with PAW pseudopotentials. However, we shall show that the error thus caused is very small in general and will not cause severe problems for the extended FT-DFT method. For detailed discussions on this issue, the reader is referred to the Appendix.

In most of our calculations, PAW pseudopotentials are employed; while unscreened Coulomb potential is used occasionally in the cases of extremely high density or high temperature. All of the potentials are generated using the \texttt{ld1.x} program included in the \texttt{Quantum-Espresso} package.

To better represent the spatially unbounded feature of electronic orbitals at high energy, a k-point mesh is used in our implementation. Previous FPMD calculations on dense hydrogen\cite{Lorenzen2010} show that $\Gamma$-point sampling of the Brillouin zone causes large fluctuations in the calculation of pressure. In our work,  a shifted 2$\times$2$\times$2 M-P k-point mesh is used to control the numerical error in both energy and pressure within 1\%.  In the  \texttt{Quantum-Espresso} package, the program fixes the maximum number $N_c$ of electronic orbitals which are explicitly calculated using the Kohn-Sham equation for each k-point. This leads to slightly different $E_c$'s among the k-points. In this case, the averaged $E_c$ is used then. 

The calculation of $U_0$  is performed in the energy interval between $E_s$ and $E_c$. The latter is the onset of the plane-wave approximation as introduced in Eq.(\ref{eq_n_rev}). The length of the interval $E_c-E_s$ is adjustable according to the demand of accuracy. In practice, $E_c$ and $E_s$ are determined from a test calculation with reduced atom number. In typical calculations, as a rule of thumb, there are 2-4 bands per electron within $E_c$, and the interval $E_c-E_s$ consists of 80-200 bands.  The effective potential $U_0$ in Eq.(~\ref{eq_dos})  is then calculated as the average of $\epsilon_i+\left<\psi_i\right|\nabla^2\left|\psi_i\right>/2$ for all orbitals in this interval, which is equivalent to the fit to the DOS in the interval but more computational economical.  The overall numerical error is less than 1\% as compared with available PIMC data of deuterium.\cite{hu2011PIMC} For high Z materials such as gold (Au) and uranium (U), the $E_c$ can be further reduced since they have more unbounded electrons and thus afford a stronger screening effect.  
Choosing $E_c$ and the energy interval in this way implies that the computational costs do not increase with temperature but have an upper bound determined by $E_c$, which makes the calculation of hot plasmas affordable with current computational resources. 

The contribution of electronic states higher than $E_c$ in
 Eqs.(\ref{eq_n_rev})-(\ref{eq_s_rev})
is numerically integrated using the Simpson method with an energy interval of 0.001 eV. The upper limit of the integration is placed at the energy of which the occupation number  is less than 10$^{-16}$ unless particularly specified.

In our calculations, the LDA and GGA functionals for the exchange-correlation interaction between electrons are employed throughout. For the LDA functional, the Perdew-Wang parameterization\cite{pw} on the Monte Carlo results of Cerperly-Alder\cite{ceperley1980} is used. For the GGA functional, the Perdew-Burke-Ernzerhof (PBE) formula\cite{pbe} is used. Also considered is the recent parameterization of finite-temperature LDA functional (FTXC) by Karasiev {\it et al.}\cite{ksdt} on the path integral Monte Carlo results of Brown {\it et al.}\cite{PIMCelectrongas}

\subsection{Numerical details on BOMD}
Thermal properties of plasmas are calculated using the ext-FPMD method together with the BO approximation in a periodic cubic calculation box. Forces on each ion are calculated through Eq.(\ref{eq_f}). Ion temperature is controlled by the Andersen thermostat,\cite{andersen1980} and the electrons are set at the same temperature as the ion, i.e., the calculation is carried out in a canonical (NVT) ensemble. Convergence with respect to  the plane-wave energy cutoff and k-point mesh is carefully examined to assure the computational error less than 1\%.  In the cases when PAW pseudopotentials are used, the core radii of the pseudopotential are carefully determined to avoid artifacts at high density and high temperature.\cite{wangcong2013}  

The time step $\Delta t$ of the ext-FPMD simulation is set to be $\Delta t =(1/40) (3/4\pi n_I)^{1/3}/\sqrt{T/m_{I}}$, i.e., roughly 1/80 of the time for an ion traveling through the average distance between two adjacent ions with the thermal velocity. Here, $n_I$ is the concentration of ions, and $m_I$ is the mass of ions.\cite{dt} In each simulation, the plasma is thermalized for 5000 MD steps to reach the equilibrium. After that, another 1000 MD steps along the ionic trajectory are used to calculate the thermal properties required.

\section{Results and discussion}

To illustrate the feature of the proposed ext-FPMD method, calculations on typical plasmas are presented as examples in this section, including aluminum (Al), deuterium (D) , helium (He),  and LiD$_2$. These materials themselves are important low Z materials of great interest to stellar physics, ICF physics, as well as laboratory astrophysics, and  have been well documented in the literature. 

\subsection{Benchmark Calculation of Hot Electrons in Al Lattice}

\begin{table}
\centering
\caption{\label{tab1}Benchmark calculation of hot electrons for $T_e$ = 30 Ry in a cold FCC Al lattice at $\rho$ = 2.7 g/cm$^3$ using the extended FT-DFT, in comparison with the usual FT-DFT calculation as the reference. Calc.~1 displays a close comparison of the extended FT-DFT results with the reference calculation,  which ignores high energy bands of occupation less than $10^{-5}$. These high energy bands are taken into consideration in Calc.~2. Calc.~3 displays the effect of a low $E_c$. Energies are presented in Ry, and pressures in Mbar.  
\\
}
\renewcommand\arraystretch{1.5}
\begin{tabular}{ccp{0.3cm}ccc}
\hline
\hline
\multirow{2}{1cm}{ }    &Reference & & \multicolumn{3}{c}{Extended  FT-DFT}  \\
        & FT-DFT  & &	 Calc.~1 &	 Calc.~2 & Calc.~3\\
 \hline 
     $\mu$             &  -128.839   & &   -128.874     &  -129.024 &  -128.992    \\

     $E$                      &-2194.664   & & -2195.021  &-2196.833 & -2196.237  \\

     $-TS$    & -2405.554  && -2406.170 & -2419.209 & -2405.237 \\

    $P$ &  410.805  && 411.885 & 421.429 & 407.532\\
\hline
\hline
\end{tabular}
\end{table}

The overall performance of the extended FT-DFT is evaluated through a benchmark calculation of hot electrons in the FCC Al lattice, which is carried out at solid density (2.7 g/cm$^3$) and electronic temperature $T_e$ = 30 Ry.  The reference calculation is performed using the usual FT-DFT method on a primitive FCC unit cell with only one Al atom. A shifted M-P 4$\times$4$\times$4 k-point mesh is used to resolve the orbitals in this small unit cell. A PAW pseudopotential similar to that in Ref.~\onlinecite{zhang2015arxiv} is used to represent the ion-electron interaction, which is nearly a Coulomb potential with a small core cutoff radius of 0.5 Bohr, and explicitly consists of all $1s$ $2s$ $2p$ $3s$ $3p$ electronic orbitals in the calculation. The plane-wave cutoff is 500 Ry. To include all electronic states with occupation number larger than 10$^{-5}$, 5000 electronic states are used in the calculation, which corresponds to an energy of 190.26 Ry. 

As a comparison, the calculations with the extended FT-DFT use only 500 orbitals explicitly, corresponding to $E_{c}$ = 39.97 Ry. These orbitals consist of 85\% of the total charge. The other 15\% is represented by plane waves. The $E_s$ is 12.35 Ry lower than $E_{c}$, and the energy bands between these two energies are used to determine $U_0$. 
In Table.~\ref{tab1}, two results calculated using the extended FT-DFT are presented. In the first calculation, denoted as Calc.~1 in the table, the contribution of the free electron in Eqs.(\ref{eq_n_rev})-(\ref{eq_s_rev}) is integrated to the occupation of 10$^{-5}$, the same as that in the usual FT-DFT calculation. As displayed in the table, Calc.~1 differs from the reference calculation less than 0.3\% for all critical parameters of plasmas, such as $\mu$, $E$, $TS$ and $P$. The comparison shows that the substitution of  electronic orbitals between 39.97 Ry and 190.26 Ry with plane waves can satisfy the accuracy requirement in the calculation of hot plasmas at much reduced computational costs.

\begin{figure}
\centering
\includegraphics[scale=0.3]{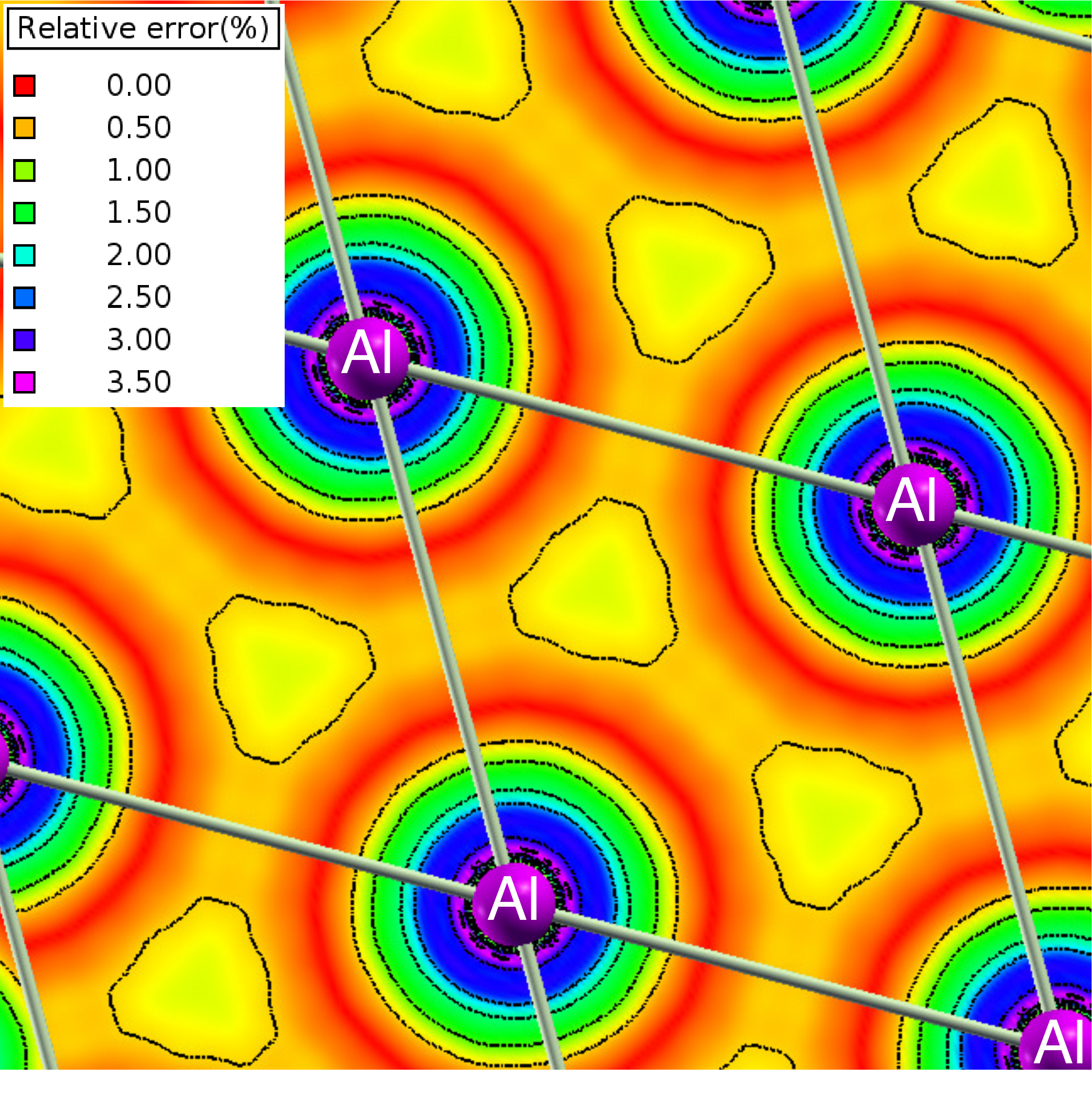} 
\caption{\label{fig5}(Color online) Relative error of total charge density on the (11$\bar{1}$) plane in the benchmark calculation using the extended FT-DFT, compared with the reference calculation using the usual FT-DFT method. The relative error is defined as $\Delta=|n({\bf r})-n^{^{\mathcal R}}({\bf r})|/n({\bf r})$.  The largest error $\sim$ 3.9\% appears in the core region.}
\end{figure}

In the second calculation, denoted as Calc.~2, the integration of free electron contribution is converged to occupation less than 10$^{-16}$.  It shows that electronic states of occupation less than 10$^{-5}$ still have considerable contribution to the calculation of $P$. Comparing with Calc.~2, the reference calculation underestimates $P$ by about 2.5\%.   
The difference of charge density between the reference calculation and Calc.~2 in Table~\ref{tab1} is displayed in Fig.~\ref{fig5}. The relative error $\Delta$, defined as $\Delta=|n({\bf r})-n^{^{\mathcal R}}({\bf r})|/n({\bf r})$, is less than 1\% overall. The largest relative error, $\sim$ 3.9\%, appears in the core region. 

In an additional calculation, Calc.~3, the effect of  low $E_c$ is probed. The $E_c$ in the calculation is 7.25 Ry. There are only 52 electronic orbitals explicitly included in the calculation, and the top 12 bands are used to calculate $U_0$, corresponding to a $E_s$ = 5.48 Ry. The occupation number is converged to 10$^{-5}$. Compared with the reference calculation, the error in the calculation of thermal parameters is still under 1\%.  However, charge density will have an overall relative error $\sim$5\%, and the largest relative error arrives at 14\% in the core region.

\subsection{Multiple-Shell Effect in Shock-Compressed Al}

The ext-FPMD method is naturally a convenient tool for the description of multiple-shell effect in hot dense plasmas, as it explicitly calculates electronic orbitals from the Kohn-Sham equation.
We use the ext-FPMD method to illustrate the multiple-shell effect in shock-compressed Al.

The calculation is carried out in a periodic cubic calculation box with 32 Al atoms,  focused on the principal Hugoniot at compression ratio $\eta\sim$ 4, where the effect of multiple shell structure is significant. A PAW pseudopotential of 0.5 Bohr core cutoff radius and 300 Ry plane-wave cutoff energy is used. All 13 electrons are explicitly included in the calculation for a better description of core electrons at high temperature. The LDA functional is used for the exchange correlation interaction, and a shifted M-P 2$\times$2$\times$2 grid is used for the integration of the first Brillouin zone. The there 960 bands explicitly included in the calculation, and the top 160 are used to calculate the value of $U_0$.
At selected thermal states, the calculation is carried out with 64 atoms to check the convergence with respect to atom number, which displays small variations on the order of 0.1\% for all quantities of interest.  

\begin{figure}
\centering
\includegraphics[scale=0.3]{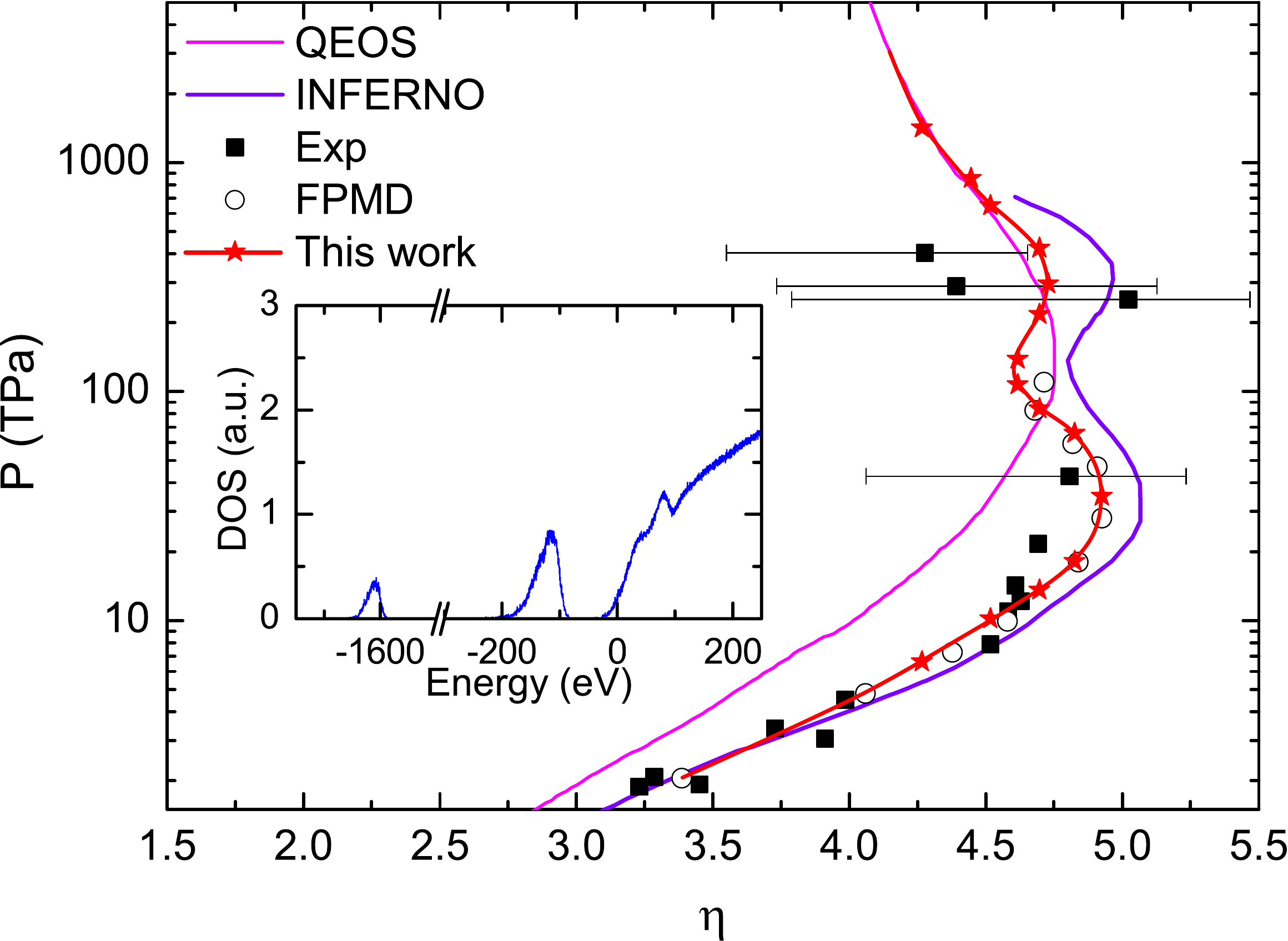} 
\caption{\label{fig10} (Color online) Principal Hugoniot of Al calculated using the ext-FPMD method with LDA for exchange-correlation interaction, compared with the FPMD calculation,\cite{surh2001} and other numerical results of the QEOS model\cite{more1988} as well as the INFERNO model.\cite{Rozsnyai2001} Experimental data are taken from Ref.~\onlinecite{surh2001, Rozsnyai2001}. The multiple-shell structure is shown as ``bumps'' around the maximum compression ratio $\eta \sim 5$. Also displayed in the inset is the DOS of Al at $\eta$ = 4.81, close to the maximum compression ratio, which shows a combination of atomic-like discrete structures at low energy and continuous spectra at high energy. }
\end{figure}

The principal Hugoniot is governed by the Rankie-Hugoniot relation
\begin{equation}\label{eq8}
E_1-E_0=\frac{\Omega_0}{2}(P_1+P_0)(1-\frac{1}{\eta}),
\end{equation}
with the subscript 0 and 1 denoting the initial state and the shocked state respectively. In addition, a quadratic polynomial interpolation is used to solve Eq.(\ref{eq8}), which reduces the numerical uncertainty to less than 0.1\%  in solving the equation. 

As displayed in Fig.~\ref{fig10}, our calculation covers the entire region of multiple-shell effect near the largest $\eta$. The upper bound of the region has a pressure more than 1.5$\times$10$^3$ TPa, which is an order higher than the maximum pressure $\sim$ 100 TPa calculated using the usual FPMD method with only 4 atoms included.\cite{surh2001} Similar to the FPMD results, the calculated Hugoniot agrees well with experimental measurements when $P <$ 100 TPa. In addition,  it goes through the center area of the experimental data\cite{Rozsnyai2001} at $P\sim$ 400 TPa and coincides with the Hugoniot of QEOS model\cite{more1988} around 650 TPa, which gives further supports to the validity of our calculation.  Fig.~\ref{fig10} also displays the multiple-bump structure near the maximum $\eta$, which is predicted by the model calculations, e.g., the INFERNO model, \cite{Rozsnyai2001} as well, and is attributed to the effect of multiple-shell structure of electrons in the K and L shells. 

In addition to the thermal properties of plasmas, the ext-FPMD method also provides the information of electronic structures, which are complicate at low energy and important for the calculation of optical properties, e.g., the X-ray absorption and opacity. The inset of Fig.~\ref{fig10} displays the DOS of the compressed Al at $\eta$ = 4.81. The atomic-like discrete structures with broadenings are clearly shown at low energy. This is an attracting feature of the ext-FPMD method, considering that the electronic structures can be used as a starting point for the calculation of ionization properties and optical properties for hot plasmas with multiple shells of electrons, which were measured in recent shock experiments.\cite{vinko2014, ciricosta2012LCLS, hoarty2013}

\subsection{Deuterium}

\begin{figure*}
\begin{minipage}{0.55 \textwidth}
\includegraphics[width=\textwidth]{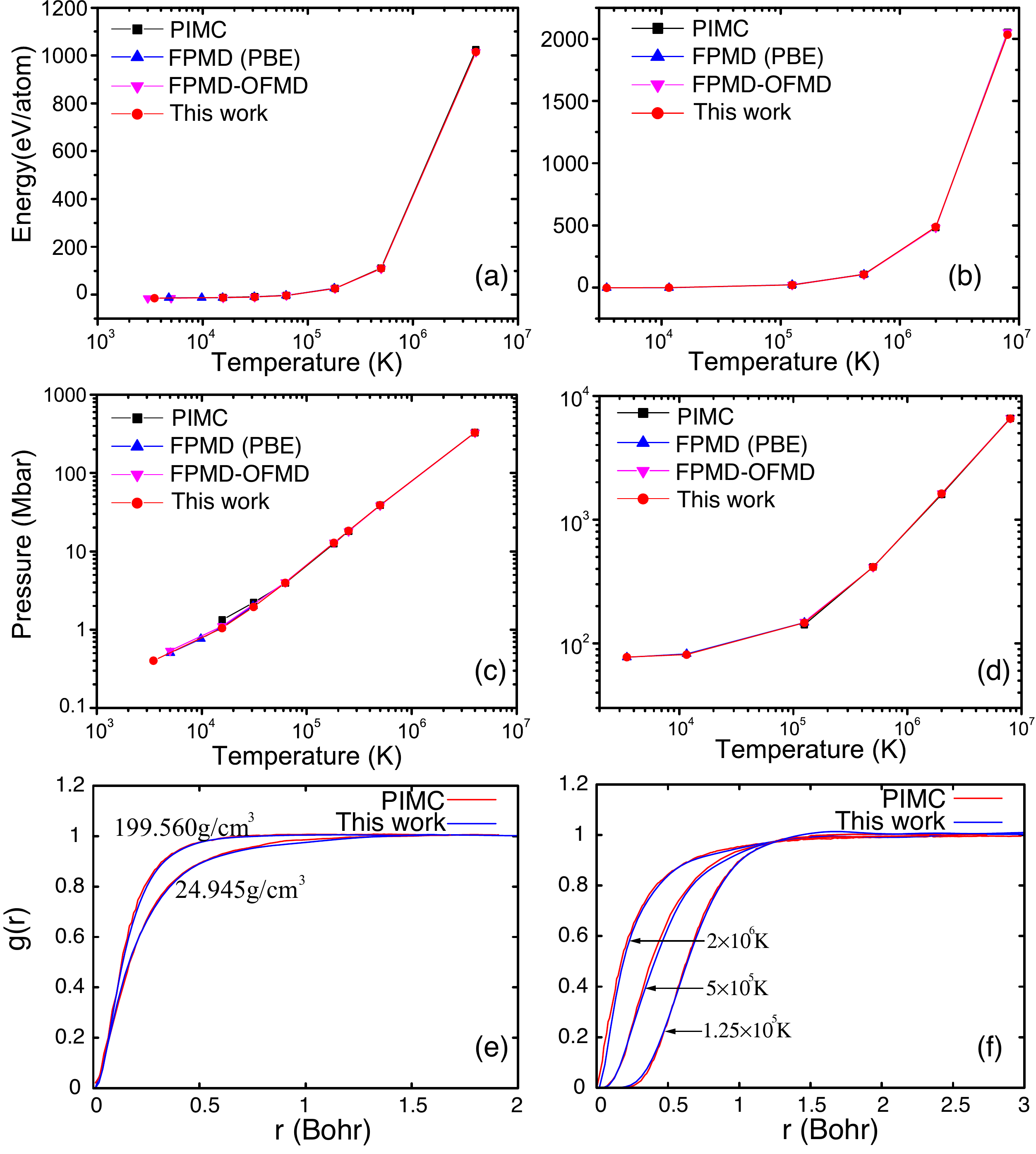} 
\end{minipage}
\begin{minipage}[c]{0.3 \textwidth}

\caption{ (Color online) Properties of warm dense  and hot dense deuterium calculated using the ext-FPMD method, compared with other first-principles methods, including PIMC,\cite{hu2011PIMC} FPMD (PBE),\cite{hu2014QMD} and FPMD-OFMD.\cite{wangcong2013} (a) Total energy per atom with respect to temperature at $\rho=1$ g/cm$^3$; (b) Total energy per atom with respect to temperature at $\rho=10$ g/cm$^3$; (c) Pressure with respect to temperature at $\rho=1$ g/cm$^3$; (d) Pressure with respect to temperature at $\rho=10$g/cm$^3$. (e) Pair distribution function $g(r)$ for selected densities at T=2$\times$10$^6$ K. (f) Pair distribution function $g(r)$ for various temperatures at $\rho=10$g/cm$^3$.  LDA\cite{pw} functional (zero-temperature) is used for all the calculations. }\label{fig6}
\end{minipage}

\end{figure*}

Deuterium is one of the well-documented plasmas because of their important role in ICF and astrophysics. Its EOS has been calculated using a number of first-principles methods including the usual FPMD methods,\cite{hu2014QMD, galli2000,lenosky2000,desjarlais2003} the OFMD method,\cite{wangcong2013} and more recently the PIMC method.\cite{hu2011PIMC, militzer2001,militzer2000}
With deuterium as an example, we show that the ext-FPMD method can be applied to a wide temperature range from $T\sim 0$ K up to $T\sim 10^{7}$ K with satisfactory accuracy for both EOS and structural properties.

The calculation is performed using a PAW pseudopotential with a core cutoff radius of 0.2 Bohr for $\rho < $ 10 g/cm$^3$, and  using the Coulomb potential with a cutoff radius of 0.002 Bohr otherwise. The plane-wave cutoff is 100 Ry - 300 Ry, which is increased with the density. 400 bands are used for the explicit Kohn-Sham calculation. The $U_0$ is calculated with the top 80 bands.
 Following the argument of  Lorenzen {\it et al.},\cite{Lorenzen2010} a shifted 2$\times$2$\times$2 M-P k-point mesh is used throughout the calculation. 
The results presented are calculated with 128 deuterium atoms. The convergence with respect to atom number are checked with 512 atoms at  $T$ = 2.5$\times$10$^5$ K and $\rho$ = 2.453 g/cm$^3$, corresponding to degeneracy parameter $\theta= T/T_F\sim$ 1. Here $T_F$ refers to the Fermi temperature. The error caused by the finite-size effect is less than 0.7\% for both $E$ and $P$, as illustrated in Table~\ref{tab_2}. Results calculated using different exchange-correlation functionals are also compared in Table~\ref{tab_2}. It shows that the difference between PBE and LDA is negligible in warm dense state at $T$ = 2.5$\times$10$^5$ K and $\rho$ = 2.453 g/cm$^3$. But there are $\sim$ 1.9\% overestimation in $P$ and $\sim$ 2.2\% in $E$ for LDA or PBE (zero-temperature) compared with the PIMC results of Hu {\it et al.}\cite{hu2011PIMC} The employment of FTXC reduces this error to less than 0.8\% with respect to the PIMC calculation. Also examined is the plasma state at $T$ = 8.0$\times$10$^6$ K and $\rho$ = 10.0 g/cm$^3$. Our calculation shows that the finite temperature correction to the exchange-correlation functionals is relatively small at high temperature. The LDA and PBE functional (zero-temperature) can converge both $P$ and $E$ within 1\% of the PIMC results.

In Fig.~\ref{fig6}(a)-(d), the $E$ and $P$ calculated using the ext-FPMD method with the LDA functional are compared with those calculated using  FPMD (with the PBE functional),\cite{hu2014QMD} OFMD,\cite{wangcong2013} and PIMC\cite{hu2011PIMC} at $\rho$ = 1.0 g/cm$^3$ and $\rho$ = 10.0 g/cm$^3$. It shows that both $E$ and $P$ agree well with the FPMD results at low temperature below 10$^4$ K. This is expected, because the occupation number above $E_c$ is essentially zero at low temperature, and the ext-FPMD method goes back to the FPMD method.
At $T >$ 10$^6$ K, our results are in line with the calculation of OFMD and PIMC, which reflects the good accuracy of our method at high temperature. The largest difference between our method and the PIMC method takes place in the calculation of $P$ at $\rho$ = 1.0 g/cm$^3$ in the temperature region between 10$^4$ K and 10$^6$ K.

\begin{table}
\centering
\caption{\label{tab_2} $E$ and $P$ calculated using the ext-FPMD method with various approximations to the exchange-correlation functions. Comparisons are made with PIMC results of Hu {\it et al.}\cite{hu2011PIMC} for warm dense deuterium at $\rho =2.4525 $ g/cm$^3$ and $T=2.5 \times 10^5$K and hot dense deuterium at $\rho$ = 10.0 g/cm$^3$ and $T$ = 8.0$\times$10$^6$ K. Calculations are performed with 128 deuterium atoms excepted those presented in parentheses, which are calculated with 512 atoms and serve as a reference for the convergence test with respect to the number of atoms.}
\renewcommand\arraystretch{1.5}

\begin{ruledtabular}
\begin{tabular}{cccp{0.3cm}cc}

 & \multicolumn{2}{c}{$\rho$ =2.4525 g/cm$^3$ }& & \multicolumn{2}{c}{$\rho =10.0 $ g/cm$^3$ } \\
&  \multicolumn{2}{c}{ $T$=2.5 $\times$ 10$^5$ K}&& \multicolumn{2}{c}{ $T$=8.0$\times$10$^6$ K} \\
 &   E(eV/atom) & P(Mbar) && E(eV/atom) & P(Mbar) \\
\hline
PIMC\cite{hu2011PIMC}& 40.55 & 45.03 &&2054.0 & 6592\\
LDA\footnotemark[1] & 41.44 & 45.89  & & 2033.7& 6563 \\
         &(41.73)\footnotemark[2] & (46.05)\footnotemark[2] && & \\
PBE\footnotemark[1] & 41.37 &45.88 &&2033.9 &6562\\
FTXC\cite{ksdt} &40.88 & 45.26 && 2042.2&6571 \\
\end{tabular}
\end{ruledtabular}
\footnotetext[1]{Zero-temperature exchange correlation}
\footnotetext[2]{Calculated with 512 atoms}

\end{table}

The pair distribution functions $g(r)$, which provide structural information of the system, are compared with the PIMC results of Hu {\it et al.}\cite{hu2011PIMC} in Fig.~\ref{fig6}(e) and (f) for $T$ and $\rho$ varying in a large range.  Very small difference is revealed between these two methods.

\subsection{$^6$LiD and Helium}

With the ext-FPMD method, we also calculate the principal Hugoniots of $^6$LiD and Helium, focusing on the warm dense region where the results of the ext-FPMD method have perceptible difference from those calculated by other first-principles methods.\cite{sheppard2014,militzer2006,militzer2009PIMC} In the calculation,  the electron-ion interactions are represented by PAW pseudopotentials. The core cutoff radius of the PAW is 0.5 Bohr. The  plane-wave cutoff are 200 Ry for deuterium and lithium, and 100 Ry for helium respectively. The k-points are resolved by a shifted 2$\times$2$\times$2 M-P mesh grid. 
In each case, the periodic cubic calculation box consists of 128 atoms, i.e., 64 formula units for $^6$LiD and 128 for He. 800 bands are explicitly included in the calculation. In the calculation of He, 1200 bands are used at selected calculations to further reduce the numerical uncertainty, which however,  only produces negligible variance to the calculation of the principal Hugoniots.  An energy interval ended at $E_c$ and consisting of 200 bands is used to estimate $U_0$ as well as the DOS of high-energy electrons. 
\begin{figure}
\centering
\includegraphics[scale=0.3]{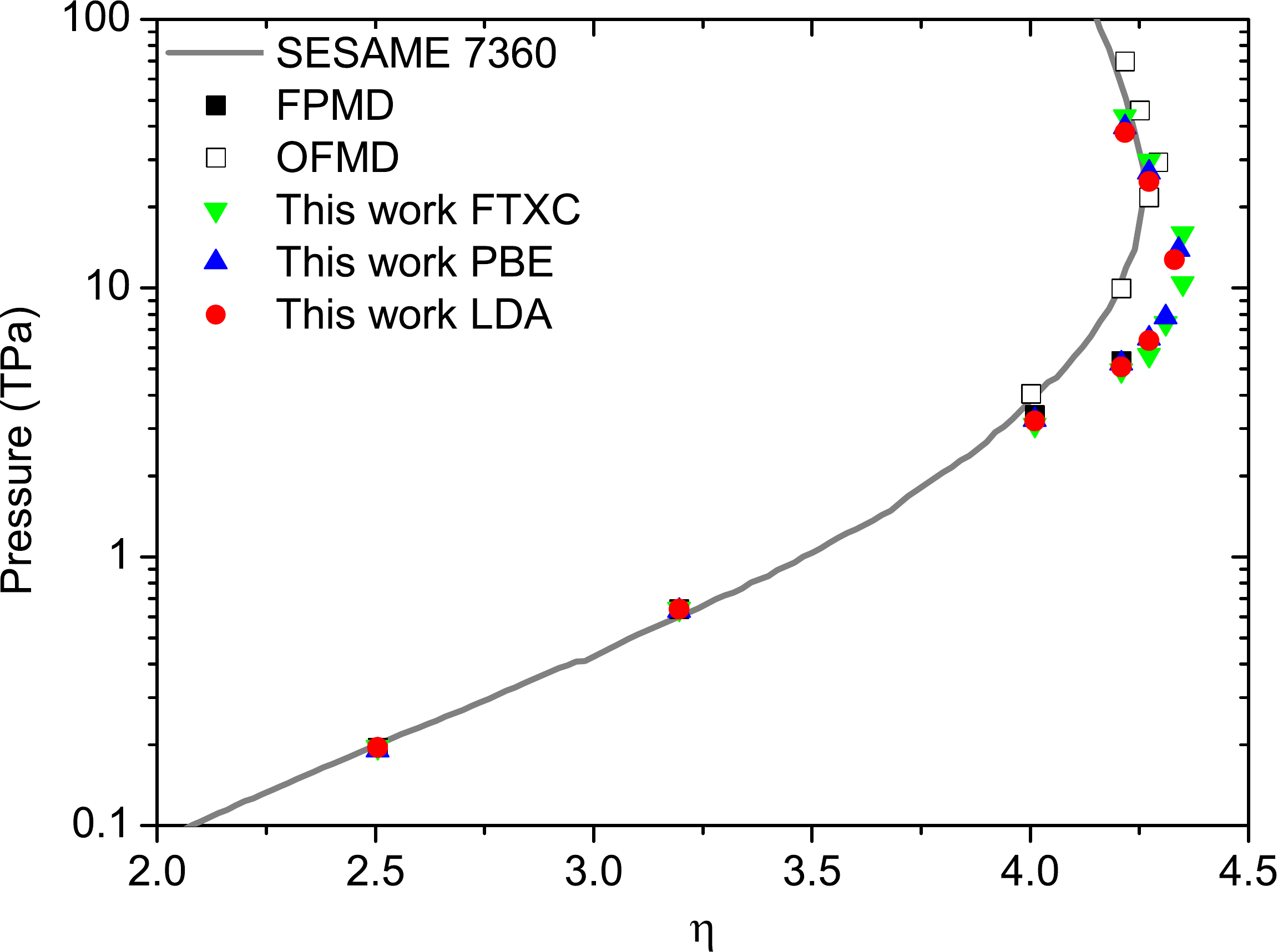} 
\caption{\label{fig9} (Color online) Principal Hugoniot of $^6$LiD calculated using the ext-FPMD method in comparison with the FPMD-OFMD results,\cite{sheppard2014} and the SESAME results.\cite{sheppard2014} Several exchange-correlation functionals are used in the calculation, including  LDA,  PBE and the FTXC\cite{ksdt}, which are labeled as solid circles, upright triangles and upside down triangles, respectively. Solid squares stand for previous FPMD results,\cite{sheppard2014} and open squares are OFMD results.\cite{sheppard2014} Notable difference is observed near the maximum compression ratio at $\eta \sim$ 4.3.
}
\end{figure}

Fig.~\ref{fig9} displays the principal Hugoniot of $^6$LiD calculated using the ext-FPMD method. 
The uncompressed state has a rock-salt crystalline structure at $\rho_0=$ 0.8g/cm$^3$. $E_0$  is calculated at $T$ = 0 K to be -110.67 eV (-108.85 eV) per formula unit for PBE (LDA and FTXC), in order to have a close comparison with  the FPMD-OFMD result of Sheppard {\it et al.}\cite{sheppard2014}
Hugoniots calculated with different exchange-correlation functionals, including LDA, PBE, and FTXC,\cite{ksdt} are distinguished by colors and symbols. It shows that the difference between exchange-correlation functionals is small below 4 TPa, which corresponds to warm dense conditions with $T<$ 10$^6$ K. However, when $P$ further increases, the results of different exchange-correlation functionals have small differences from each other, which qualitatively agrees with the trend illustrated in the calculation of Al and deuterium in warm dense states.

As a comparison, also displayed are the results calculated by the combination of FPMD and OFMD,\cite{sheppard2014} denoted as squares in the figure. The two parts of Hugoniot are connected to each other by the ``bootstrapping'' process at selected compression ratios.\cite{sheppard2014} 
The Hugoniot calculated using the ext-FPMD method is notably different from that calculated using the OFMD method near the largest compression ratio $\eta\sim$ 4.3. 
The pressure in our calculation is systematically lower than that calculated with the OFMD method up to 40\% at some compression ratios. The shape of Hugoniots and the maximum compression ratio are also different. 

\begin{figure}
\centering
\includegraphics[scale=0.3]{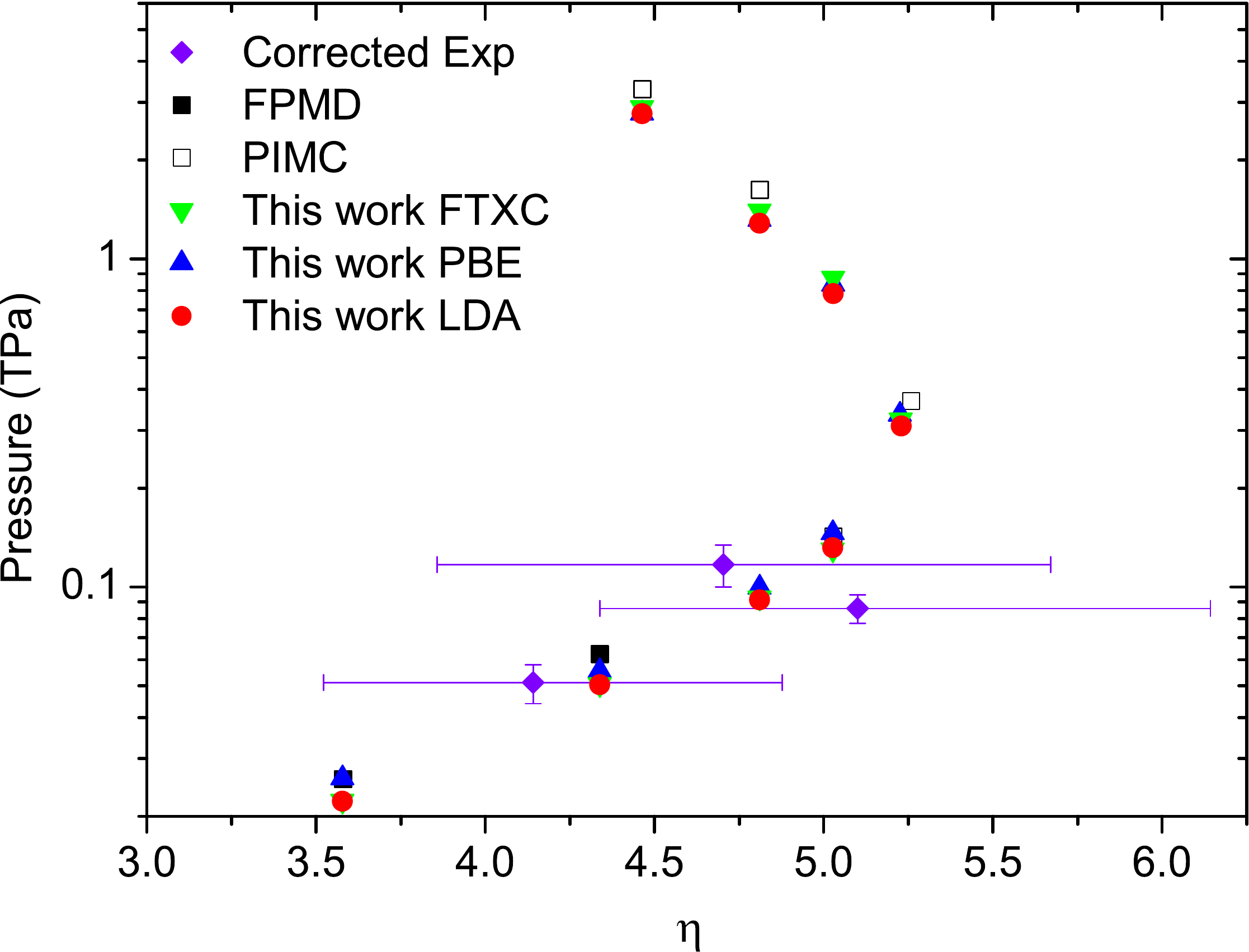} 
\caption{\label{fig8} (Color online) Principal Hugoniot of helium calculated using the ext-FPMD method. Also displayed are FPMD-PIMC results.\cite{militzer2006,militzer2009PIMC} Several exchange-correlation functionals are used in the calculation, including  LDA,  PBE and the FTXC\cite{ksdt}, which are labeled as solid circles, upright triangles and upside down triangles, respectively. Solid squares stand for previous FPMD results,\cite{militzer2006,militzer2009PIMC} and open squares are PIMC results.\cite{militzer2006,militzer2009PIMC} Experimental data shown as solid squares with error bars are taken from Ref.~\onlinecite{eggert2008} and corrected following Ref.~\onlinecite{knudson2009, celliers2010}.
}
\end{figure}

A further calculation on the principal Hugoniot of He is performed, compared with the PIMC results\cite{militzer2006, militzer2009PIMC} at high temperature and the FPMD results\cite{militzer2006, militzer2009PIMC} at low temperature. The initial state of the uncompressed helium is a liquid at $\rho_0$ = 0.1235 g/cm$^3$. 
Our calculation shows that the principal Hugoniot is not sensitive to the choice of exchange correlation functionals, but relies on the accuracy of $E_0$. In the current work, it is calculated using the FPMD method at 4.3 K to be $E_0$ = -78.72 eV/atom for PBE, and -77.13 eV/atom for LDA as well as FTXC.
Fig.~\ref{fig8} shows that Hugoniots calculated with these $E_0$ have a good agreement with the result of PIMC. The small underestimation of the ext-FPMD calculation at high pressure is attributed to the overestimation of $E_0$, e.g., $\sim$ 1.04 eV/atom for PBE as estimated by Militzer, \cite{militzer2009PIMC} combined with a different interpolation in solving Eq.(\ref{eq8}).\cite{militzer2009PIMC}

\section{Summary}
We show that the FPMD method can be readily extended to high temperature region of dense plasmas with a satisfactory accuracy when the plane-wave approximation at high energy is appropriately integrated. The extended method not only provides a systematic way to calculate thermal properties, e.g., pressure, internal energy, and free energy, from cold materials to hot dense plasmas, it can also be used as a good starting point for the theoretical investigation of the lowering of ionization potential, the X-ray absorption/emission spectra, opacity , and highly ionized high-Z materials, which are of great interest to high energy density physics, ICF, and laboratory astrophysics.

\begin{acknowledgments}
This work is financially supported by the NSFC (Grant No. 11274019) and NSFC-NSAF (Grant No. U1530113). 
\end{acknowledgments}

\appendix*
\section{Influence of  ${\hat V}_l^{NL}$ on the plane-wave approximation at high energy}
It is convenient to write ${\hat V}^{NL}_l$ in a semi-local form as
$
{\hat V}^{NL}_l=\sum_m\left|Y_{lm}\right>V_l(r)\left<Y_{lm}\right|,
$
where $Y_{lm}$ is the spherical harmonics. $V_l(r)$ is a finite correction at the vicinity of each ion within the radius $r_c$, and it vanishes outside.  At high energy, in addition to the expression of Eq.(\ref{eq_e_rev}) for the total energy $E^{^{\mathcal R}}$, the operator has an energy-dependent contribution proportional to
$$
{\mathcal G}_l({\bf k})=\sum_{I,m} \int d{\bf r}\left<{\bf k}\right|\left.Y_{lm}\right>V_l(|{\bf r}-{\bf R}_I|)\left<Y_{lm}\right.\left|{\bf k}\right>,
$$
when the plane-wave approximation is employed. Here, $\bf k$ is the wave vector  and $k^2/2+U_0=\epsilon$, and the plane wave is represented by $\left|{\bf k}\right>=\Omega^{-1/2}\exp(-i{\bf k}\cdot{\bf r})$.  

Using the spherical Bessel functions $j_l$, ${\mathcal G}_l({\bf k})$ can be further expressed as\cite{nielsen1985a}
$$
{\mathcal G}_l({\bf k})=\frac{4\pi N_I}{\Omega}(2l+1)\int_0^{r_c}dr j_l^2(kr)V_l( r)r^2,
$$
with $N_I$ the number of ions. The integration can be carried out approximately at high energy, i.e., $k\gg1$, as 
$$
{\mathcal G}_l({\bf k})\approx \frac{2\pi N_I }{\Omega}(2l+1) \frac{r_c V_l(0)}{k^2}=\frac{C_l}{\Omega}\frac{1}{k^2},
$$
noticing that $\int dx x^2 j_l^2(x)=(x^3/2)(j^2_{l}-j_{l-1}j_{l+1})$, and $j_l(x)$ ($\sim \sin(x-l\pi/2)/x$ at large $x$) is strongly oscillating. \cite{arfkenbook}

Then, the overall contribution of  all $k> \sqrt{2(E_c-U_0)}$ to $E^{^{\mathcal R}}$  is
\begin{equation}
\begin{aligned}
E^{^{NL}}&\approx \sum_{l}\frac{C_l}{\Omega} \int_{E_c}^{\infty} d\epsilon  \frac{D(\epsilon) f(\epsilon)}{\epsilon-U_0} \\
               &\approx \sum_l  \sqrt{\frac{C_l^2 T}{2\pi^3}}e^{(\mu-U_0)/T}\erfc\left(\sqrt{\frac{E_c-U_0}{T}}\right),
\end{aligned}
\end{equation}
where $\erfc(x)$ is the complementary error function. The magnitude of $E^{^{NL}}$ is controlled by $e^{(\mu-U_0)/T}$, which approaches  zero as $T^{-3/2}$. This is evident because $\mu=T\ln\left({n_e}/{n_Q}\right)$ for free electrons at high temperature,\cite{kittel_tp} where $n_e$ is the concentration of electrons, and $n_Q$ is the quantum concentration defined as $n_Q=\left(T/2\pi\right)^{3/2}$.

Since $\ E^{^{NL}}$ is independent to ${\bf R}_I$, the non-local operator does not contribute an extra correction to ${\bf F}_I$, which is now calculated through Eq.(\ref{eq_f}).  Corresponding correction to the stress term $\sigma_{\alpha\beta}^{ei}$ can be obtained by applying an infinitesimal strain $r_\alpha \rightarrow r_\alpha+\epsilon_{\alpha\beta}r_\beta$ to $E^{^{NL}}$ under isentropic conditions, which gives
$$
\sigma^{ei,NL}_{\alpha\beta}=-\frac{\delta_{\alpha\beta}}{3\Omega}E^{^{NL}}.
$$
Note that $T\Omega^{2/3}$ is a constant for free electrons under isentropic conditions, and $\Omega\rightarrow(1+\epsilon_{11}+\epsilon_{22}+\epsilon_{33})\Omega$ under the strain. This term decreases the same as $E^{^{NL}}$ with respect to $T$. In addition, 
since $\erfc\left(\sqrt{\frac{E_c-U_0}{T}}\right)$ approaches to zero quickly when $E_c$ increases, the residue error is even smaller for a large $E_c$.  So, in principle, there is no fundamental difficulty to use NCPP with the plane-wave approximation at high energy.  The cost is a slightly reduced accuracy, or  increased computational expenses. 

The ultrasoft psuedopotential has a similar (more specifically linearized) cancellation process for the non-local operator ${\hat V}^{NL}_l$ as the PAW pseudopotentials.\cite{kresse1999} But it is usually built on a reference NCPP because of its difficulty in calculating the compensating charge in the plane-wave bases.\cite{kresse1999} Its non-locality in $V^{ei}$ is actually determined by the underlying reference NCPP, and so does the residue error.

\bibliographystyle{apsrev4-1}
\bibliography{free_electron}
\end{document}